\def\year{2020}\relax
\documentclass[letterpaper]{article} 
\usepackage{aaai20}  
\usepackage{times}  
\usepackage{helvet} 
\usepackage{courier}  
\usepackage[hyphens]{url}  
\usepackage{graphicx} 
\usepackage{amsmath}
\usepackage{booktabs}
\usepackage{algorithm}
\usepackage{algorithmic}
\usepackage{amsfonts}
\usepackage{subfigure}
\usepackage{multirow}
\usepackage{array}
\usepackage{graphicx}  
\title{GMAN: A Graph Multi-Attention Network for Traffic Prediction}
\author{Chuanpan Zheng\textsuperscript{\rm 1,2,3}, Xiaoliang Fan\textsuperscript{\rm 1,2,3}\thanks{Corresponding author}, Cheng Wang\textsuperscript{\rm 1,2,3}, Jianzhong Qi\textsuperscript{\rm 4}\\ 
\textsuperscript{\rm 1}Fujian Key Laboratory of Sensing and Computing for Smart Cities, Xiamen University, Xiamen, China\\
\textsuperscript{\rm 2}Digital Fujian Institute of Urban Traffic Big Data Research, Xiamen University, Xiamen, China\\ 
\textsuperscript{\rm 3}School of Informatics, Xiamen University, Xiamen, China\\ 
\textsuperscript{\rm 4}School of Computing and Information Systems, University of Melbourne, Melbourne, Australia\\
zhengchuanpan@stu.xmu.edu.cn,
\{fanxiaoliang, cwang\}@xmu.edu.cn,
jianzhong.qi@unimelb.edu.au 
}

\begin{document}
	
\maketitle

\begin{abstract}
	Long-term traffic prediction is highly challenging due to the complexity of traffic systems and the constantly changing nature of many impacting factors. In this paper, we focus on the spatio-temporal factors, and propose a graph multi-attention network (GMAN) to predict traffic conditions for time steps ahead at different locations on a road network graph. GMAN adapts an encoder-decoder architecture, where both the encoder and the decoder consist of multiple spatio-temporal attention blocks to model the impact of the spatio-temporal factors on traffic conditions. The encoder encodes the input traffic features and the decoder predicts the output sequence. Between the encoder and the decoder, a transform attention layer is applied to convert the encoded traffic features to generate the sequence representations of future time steps as the input of the decoder. The transform attention mechanism models the direct relationships between historical and future time steps that helps to alleviate the error propagation problem among prediction time steps. Experimental results on two real-world traffic prediction tasks (i.e., traffic volume prediction and traffic speed prediction) demonstrate the superiority of GMAN. In particular, in the 1 hour ahead prediction, GMAN outperforms state-of-the-art methods by up to 4\% improvement in MAE measure. The source code is available at \url{https://github.com/zhengchuanpan/GMAN}.
\end{abstract}

\section{Introduction}

Traffic prediction aims to predict the future traffic conditions (e.g., traffic volume or speed) in road networks based on historical observations (e.g., recorded via sensors). It plays a significant role in many real-world applications. For example, the accurate traffic prediction can help transportation agencies better control the traffic to reduce traffic congestion~\cite{Lv-et-al:IJCAI2018,Zheng-et-al:TITS2019}.

The traffic conditions at nearby locations are expected to impact each other. To capture such spatial correlations,~\textit{Convolutional neural networks} (CNN) are widely used~\cite{Zhang-et-al:AAAI2017,Yao-et-al:AAAI2018,Yao-et-al:AAAI2019}. Meanwhile, The traffic condition at a location is also correlated with its historical observations.~\textit{Recurrent neural networks} (RNN) are widely applied to model such temporal correlations~\cite{Ma-et-al:TRC2015,Song-et-al:IJCAI2016}.            

Recent studies formulate the traffic prediction as a graph modeling problem, since the traffic conditions are restricted on road network graphs~\cite{Li-et-al:ICLR2018,Yu-et-al:IJCAI2018,Wu-et-al:IJCAI2019}. Using~\textit{graph convolutional networks} (GCN)~\cite{Defferrard-et-al:NIPS2016}, these studies achieve promising results for short-term (5\url{~}15 minutes ahead) traffic prediction. However, the long-term (up to a few hours ahead~\cite{Hou-and-Li:TITS2016}) traffic prediction still lacks a satisfactory progress in the literature, mainly due to the following challenges.

\begin{figure}
	\centering
	\subfigure[]{
		\label{Figure1(a)} 
		\includegraphics[width = 0.19 \columnwidth]{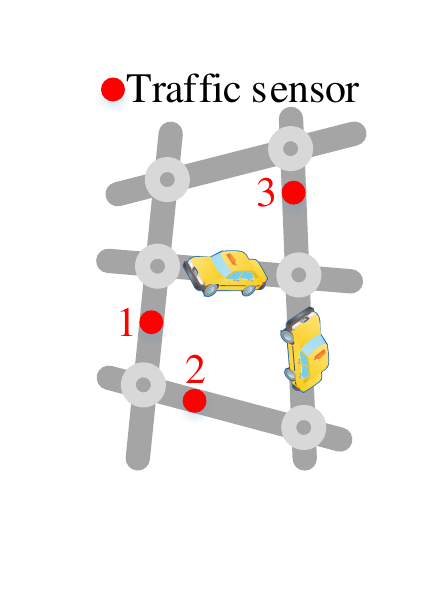}}
	\hspace{0.0 \columnwidth}
	\subfigure[]{
		\label{Figure1(b)} 
		\includegraphics[width = 0.67 \columnwidth]{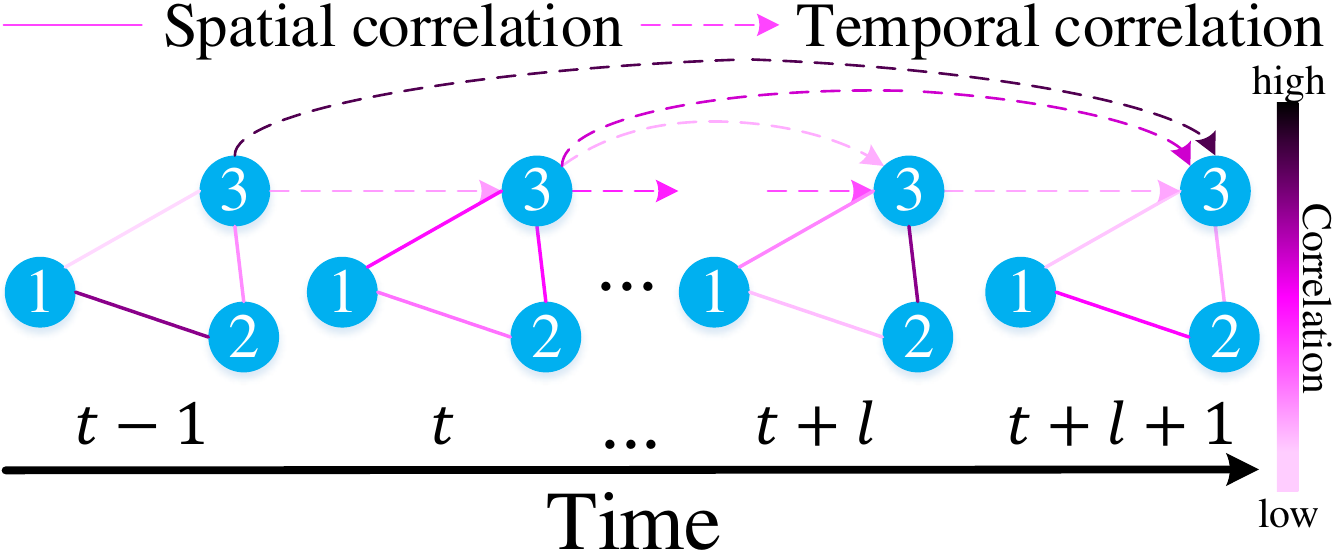}}
	\caption{Complex spatio-temporal correlations. (a) Sensors in a road network. (b) Dynamic spatial correlations: sensors 1 and 2 are not always highly correlated, although they are close in the road network; non-linear temporal correlations: the traffic condition of sensor 3 at time step $ t+l+1 $ may be more correlated to that of distant time steps (e.g., $ t-1 $) rather than recent time steps (e.g., $ t+l $).}
	\label{Figure1} 
\end{figure}

\noindent 1) \textit{Complex spatio-temporal correlations.}
\begin{itemize}
	\item Dynamic spatial correlations. As shown in Figure~\ref{Figure1}, the correlations of traffic conditions among sensors in a road network change significantly over time (e.g., before and during peak hours). How to dynamically select relevant sensors' data to predict a target sensor's traffic conditions in long-term horizon is a challenging issue.
	\item Non-linear temporal correlations. Also in Figure~\ref{Figure1}, the traffic condition at a sensor may fluctuate tremendously and suddenly (e.g., because of an accident), affecting the correlations between different time steps. How to adaptively model the non-linear temporal correlations when the time goes further into the future remains a challenge.
\end{itemize}

\noindent 2) \textit{Sensitivity to error propagation.} In the long-term horizon, small errors in each time step may amplify when predictions are made further into the future. Such error propagations make predictions into far future highly challenging.

To address the aforementioned challenges, we propose a \textit{\underline{G}raph \underline{M}ulti-\underline{A}ttention \underline{N}etwork} (GMAN) to predict traffic conditions on a road network graph over time steps ahead. Here, the traffic conditions refer to observations over a traffic system that can be reported in numeric values. For illustration purpose, we focus on traffic volume and traffic speed predictions, although our model could be applied to predictions of other numerical traffic data.

GMAN follows the encoder-decoder architecture, where the encoder encodes the input traffic features and the decoder predicts the output sequence. A transform attention layer is added between the encoder and the decoder to convert the encoded historical traffic features to generate future representations. Both the encoder and the decoder are composed of a stack of \textit{ST-Attention blocks}. Each ST-Attention block is formed by a spatial attention mechanism to model the dynamic spatial correlations, a temporal attention mechanism to model the non-linear temporal correlations, and a gated fusion mechanism to adaptively fuse the spatial and temporal representations. The transform attention mechanism models direct relationships between historical and future time steps to alleviate the effect of error propagation. Experiments on two real-world datasets confirm that GMAN achieves state-of-the-art performances. 

The contributions of this work are summarized as follow:
\begin{itemize}
	\item We propose spatial and temporal attention mechanisms to model the dynamic spatial and non-linear temporal correlations, respectively. Moreover, we design a gated fusion to adaptively fuse the information extracted by spatial and temporal attention mechanisms.
	\item We propose a transform attention mechanism to transform the historical traffic features to future representations. This attention mechanism models direct relationships between historical and future time steps to alleviate the problem of error propagation.
	\item We evaluate our graph multi-attention network (GMAN) on two real-world traffic datasets, and observe 4\% improvement and superior fault-tolerance ability over state-of-the-art baseline methods in 1 hour ahead prediction.
\end{itemize}

\section{Related Work}

\paragraph{Traffic Prediction} Traffic prediction has been extensively studied in past decades. Deep learning approaches (e.g., long short-term memory (LSTM)~\cite{Ma-et-al:TRC2015}) show more superior performance in capturing temporal correlations in traffic conditions, compared with traditional time-series methods (e.g., auto-regressive integrated moving average (ARIMA)~\cite{ARIMA:1997}) and machine learning models (e.g., support vector regression (SVR)~\cite{Wu-et-al:TITS2004}, k-nearest neighbor (KNN)~\cite{Zheng-et-al:TRC2014}). To model spatial correlations, researchers apply convolutional neural networks (CNN) to capture the dependencies in Euclidean space~\cite{Zhang-et-al:AAAI2017,Yao-et-al:AAAI2018,Yao-et-al:AAAI2019}. Recent studies formulate the traffic prediction on graphs and employ graph convolutional networks (GCN) to model the non-Euclidean correlations in the road network~\cite{Li-et-al:ICLR2018,Lv-et-al:IJCAI2018}. These graph-based models generate multiple steps ahead predictions via a step-by-step approach and may suffer from error propagation between different prediction steps.

\paragraph{Deep Learning on Graphs} Generalizing neural networks to graph-structured data is an emerging topic~\cite{Bronstein-et-al:SPM2017,Wu-et-al:arXiv2019}. A line of studies generalize CNN to model arbitrary graphs on spectral~\cite{Defferrard-et-al:NIPS2016,Kipf-and-Welling:ICLR2017,Li-et-al:AAAI2018} or spatial~\cite{Atwood-and-Towsley:NIPS2016,Hamilton-et-al:NIPS2017,Chen-et-al:ICLR2018} perspective. Another line of studies focus on graph embedding, which learns low-dimensional representations for vertices that preserve the graph structure information~\cite{Grover-and-Leskovec:KDD2016,Cui-et-al:TKDE2019}.~\cite{Wu-et-al:IJCAI2019} integrates WaveNet~\cite{Oord-et-al:arXiv2016} into GCN for spatio-temporal modeling. As it learns static adjacency matrices, this method faces difficulties in capturing dynamic spatial correlations.

\paragraph{Attention Mechanism} Attention mechanisms have been widely applied to various domains due to their high efficiency and flexibility in modeling dependencies~\cite{Vaswani-et-al:NIPS2017,Shen-et-al:AAAI2018,Du-et-al:arXiv2018}. The core idea of attention mechanisms is to adaptively focus on the most relevant features according to the input data~\cite{Cheng-et-al:AAAI2018}. Recently, researchers apply attention mechanisms to graph-structured data~\cite{Velickovic-et-al:ICLR2018} to model spatial correlations for graph classification. We extend the attention mechanism to graph spatio-temporal data prediction.

\section{Preliminaries}

We denote a road network as a weighted directed graph $ \mathcal{G=(V,E,A)} $. Here, $ \mathcal{V} $ is a set of $ N = | \mathcal{V} | $ vertices representing points (e.g., traffic sensors) on the road network; $ \mathcal{E} $ is a set of edges representing the connectivity among vertices; and $ \mathcal{A} \in \mathbb{R}^{N \times N} $ is the weighted adjacency matrix, where $ \mathcal{A}_{v_i,v_j} $ represents the proximity (measured by the road network distance) between vertex $ v_i $ and $ v_j $. 

The traffic condition at time step $ t $ is represented as a graph signal $ X_{t} \in \mathbb{R}^{N \times C} $ on graph $ \mathcal{G} $, where $ C $ is the number of traffic conditions of interest (e.g., traffic volume, traffic speed, etc.). 

\paragraph{Problem Studied} Given the observations at $ N $ vertices of historical $ P $ time steps $ \mathcal{X} = (X_{t_{1}},X_{t_{2}},...,X_{t_{P}}) \in \mathbb{R}^{P \times N \times C} $, we aim to predict the traffic conditions of the next $ Q $ time steps for all vertices, denoted as $ \hat{Y}=(\hat{X}_{t_{P+1}},\hat{X}_{t_{P+2}},...,\hat{X}_{t_{P+Q}}) \in \mathbb{R}^{Q \times N \times C} $. 

\section{Graph Multi-Attention Network}

\begin{figure}
	\centering
	\begin{minipage}{0.514 \columnwidth}
		\subfigure[The architecture of GMAN]{
			\label{Figure2(a)} 
			\includegraphics[width = 0.98 \columnwidth]{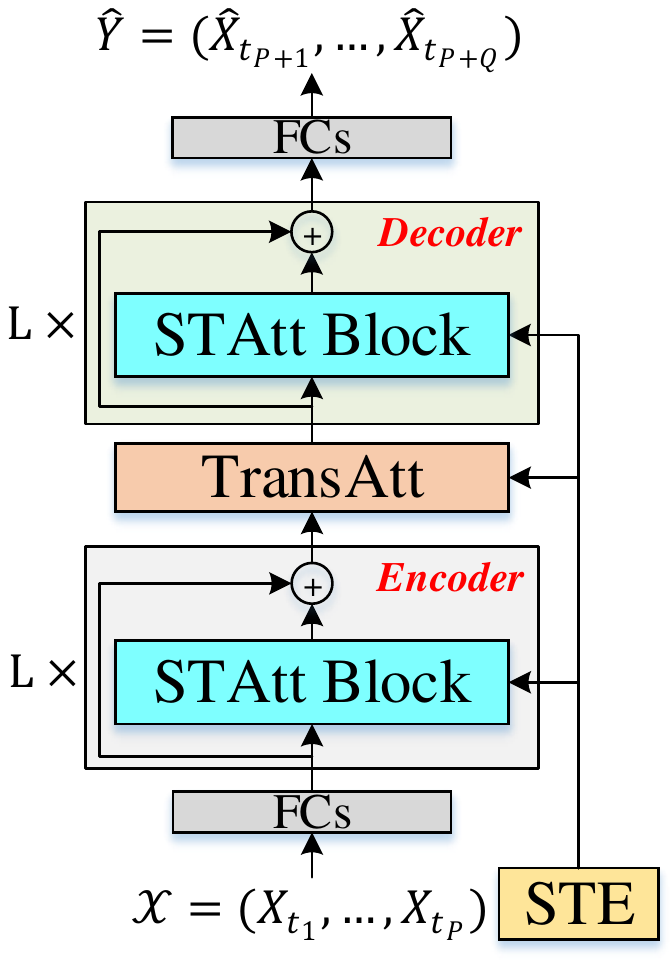}}
	\end{minipage}	
	\begin{minipage}{0.476 \columnwidth}
		\subfigure[Spatio-Tempoal Embedding]{
			\label{Figure2(b)} 
			\includegraphics[width = 0.98 \columnwidth]{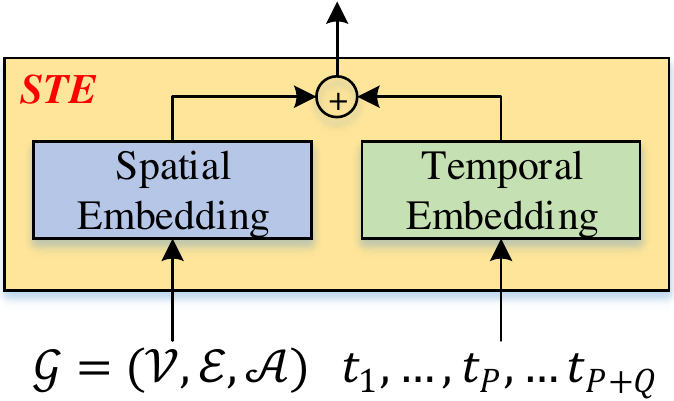}}\\
		\subfigure[ST-Attention Block]{
			\label{Figure2(c)} 
			\includegraphics[width = 0.98 \columnwidth]{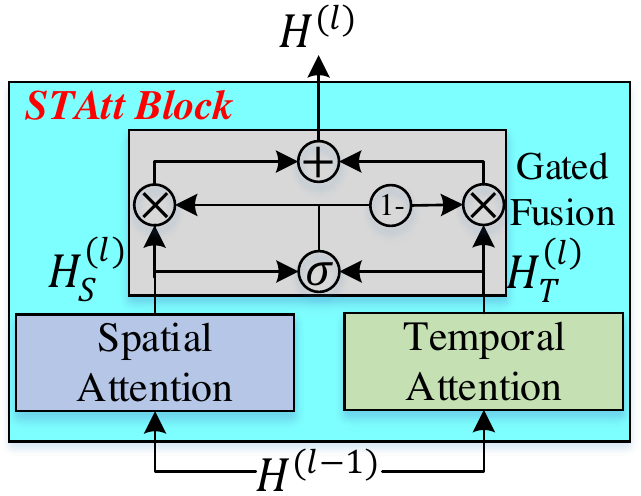}}
	\end{minipage}
	\caption{The framework of Graph multi-attention network (GMAN). (a) GMAN consists of a spatio-temporal embedding (STE), an encoder and a decoder both with $ L $ ST-Attention blocks (STAtt Block), a transform attention layer (TransAtt), and two fully-connected layers (FCs). (b) The spatio-temporal embedding contains a spatial embedding and a temporal embedding. (c) The ST-Attention block combines spatial and temporal attention mechanisms via gated fusion.}
	\label{Figure2}
\end{figure}

Figure~\ref{Figure2} illustrates the framework of our proposed \textit{graph multi-attention network} (GMAN), which has an encoder-decoder structure. Both the encoder and the decoder contain $ L $ ST-Attention blocks (STAtt Block) with residual connections~\cite{He-et-al:CVPR2016}. Each ST-Attention block is composed of spatial and temporal attention mechanisms with gated fusion. Between the encoder and the decoder, a transform attention layer is added to the network to convert the encoded traffic features to the decoder. We also incorporate the graph structure and time information into multi-attention mechanisms through a \textit{spatio-temporal embedding} (STE). In addition, to facilitate the residual connection, all layers produce outputs of $ D $ dimensions. The modules are detailed next.

\subsection{Spatio-Temporal Embedding}

Since the evolution of traffic conditions is restricted by the underlying road network~\cite{Lv-et-al:IJCAI2018}, it is crucial to incorporate the road network information into prediction models. To this end, we propose a \textit{spatial embedding} to encode vertices into vectors that preserve the graph structure information. Specifically, we leverage the \textit{node2vec} approach~\cite{Grover-and-Leskovec:KDD2016} to learn the vertex representations. In addition, to co-train the pre-learned vectors with the whole model, these vectors are fed into a two-layer fully-connected neural network. Then, we obtain the spatial embedding, represented as $ e^{S}_{v_i} \in \mathbb{R}^D $, where $ v_i \in \mathcal{V} $.

The spatial embedding only provides static representations, which could not represent the dynamic correlations among traffic sensors in the road network. We thus further propose a \textit{temporal embedding} to encode each time step into a vector. Specifically, let a day be with $ T $ time steps. We encode the day-of-week and time-of-day of each time step into $ \mathbb{R}^{7} $ and $ \mathbb{R}^{T} $ using one-hot coding, and concatenate them into a vector $ \mathbb{R}^{T+7} $. Next, we apply a two-layer fully-connected neural network to transform the time feature to a vector $ \mathbb{R}^{D} $. In our model, we embed time features for both historical $ P $ and future $ Q $ time steps, represented as $ e^{T}_{t_j} \in \mathbb{R}^D $, where $ t_j=t_1,...,t_P,...,t_{P+Q} $.

To obtain the time-variant vertex representations, we fuse the aforementioned spatial embedding and temporal embedding as spatio-temporal embedding (STE), as shown in Figure \ref{Figure2(b)}. Specifically, for vertex $ v_i $ at time step $ t_j $, the STE is defined as $ e_{v_i, t_j} = e^S_{v_{i}}  + e^T_{t_{j}} $. Therefore, the STE of $ N $ vertices in $ P + Q $ time steps is represented as $ E \in \mathbb{R}^{ (P + Q) \times N \times D} $. The STE contains both graph structure and time information, and it will be used in spatial, temporal and transform attention mechanisms. 

\subsection{ST-Attention Block}

As shown in Figure \ref{Figure2(c)}, the ST-Attention block includes a spatial attention, a temporal attention and a gated fusion. We denote the input of the $ l^{th} $ block as $ H^{(l-1)} $, where the hidden state of vertex $ v_i $ at time step $ t_j $ is represented as $ h_{v_i,t_j}^{(l-1)} $. The outputs of spatial and temporal attention mechanisms in the $ l^{th} $ block are represented as $ H_S^{(l)} $ and $ H_T^{(l)} $, where the hidden states of vertex $ v_i $ at time step $ t_j $ are denoted as $ hs_{v_i,t_j}^{(l)} $ and $ ht_{v_i,t_j}^{(l)} $, respectively. After the gated fusion, we obtain the output of the $ l^{th} $ block, represented as $ H^{(l)} $.

For illustration purpose, we denote a non-linear transformation as:
\begin{equation}
f(x) = \mathrm{ReLU} ( x \mathbf{W} + \mathbf{b} ) 
\label{non-linear transform},
\end{equation}
where $ \mathbf{W} $, $ \mathbf{b} $ are learnable parameters, and ReLU~\cite{Nair-and-Hinton:ICML2010} is the activation function.

\subsubsection{Spatial Attention}

\begin{figure}
	\centering
	\includegraphics[width = 0.90 \columnwidth]{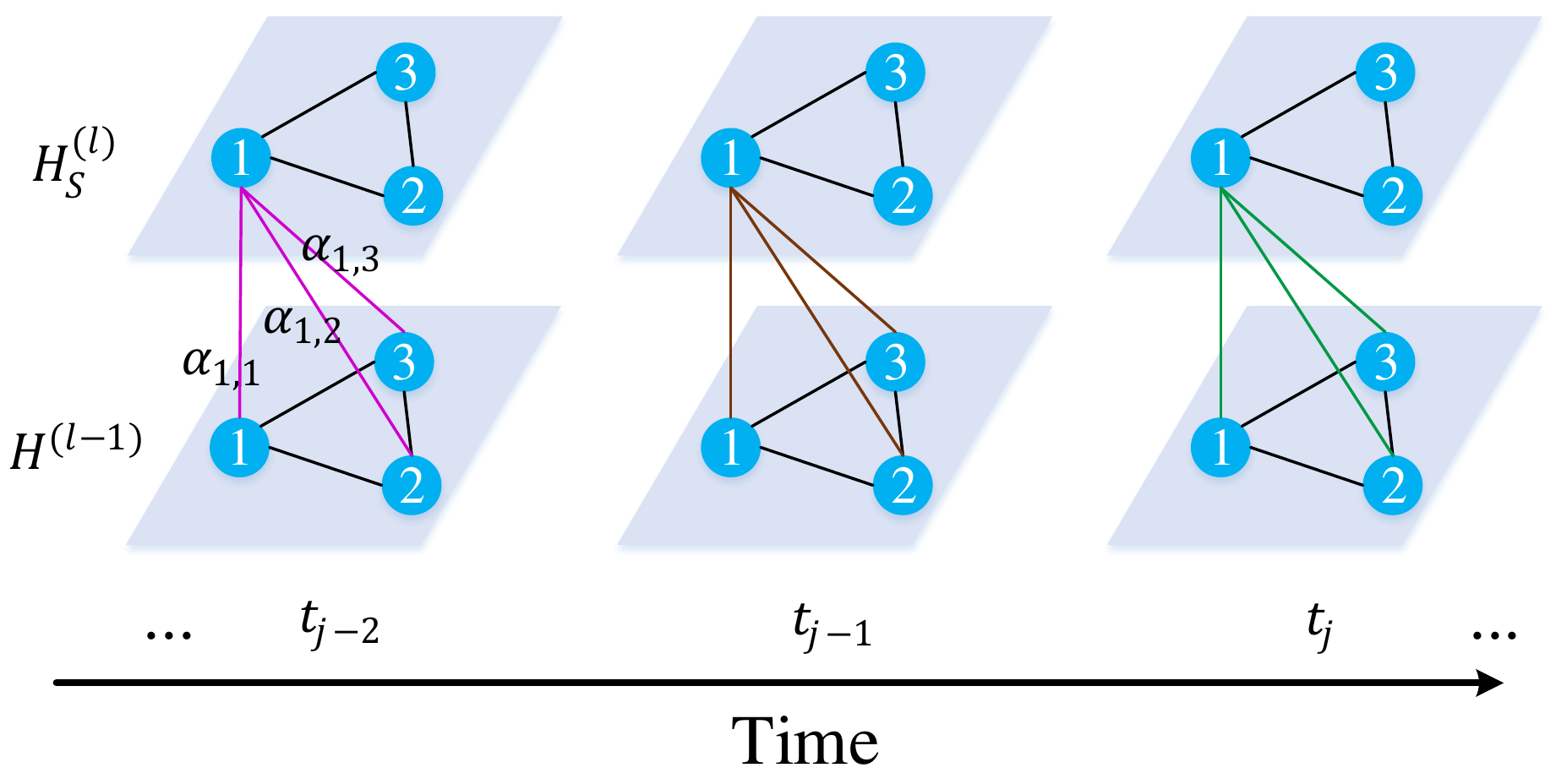} \\
	\caption{The spatial attention mechanism captures time-variant pair-wise correlations between vertices.}
	\label{Figure3}
\end{figure}

The traffic condition of a road is affected by other roads with different impacts. Such impact is highly dynamic, changing over time. To model these properties, we design a spatial attention mechanism to adaptively capture the correlations between sensors in the road network. The key idea is to dynamically assign different weights to different vertices (e.g., sensors) at different time steps, as shown in Figure~\ref{Figure3}. For vertex $ v_i $ at time step $ t_j $, we compute a weighted sum from all vertices:
\begin{equation}
hs_{v_i,t_j}^{(l)} = \sum\nolimits_{v \in \mathcal{V}} \alpha_{v_i,v} \cdot h_{v,t_j}^{(l-1)} \label{spatial attention3},
\end{equation}
where $ \mathcal{V} $ denotes a set of all vertices, $ \alpha_{v_i,v} $ is the attention score indicating the importance of vertex $ v $ to $ v_i $, and the summation of attention scores equals to $ 1 $: $ \sum\nolimits_{v \in \mathcal{V}} \alpha_{v_i,v} = 1 $. 

At a certain time step, both the current traffic conditions and the road network structure could affect the correlations between sensors. For example, a congestion on a road may significantly affect the traffic conditions of its adjacent roads. Motivated by this intuition, we consider both traffic features and the graph structure to learn the attention score. Specifically, we concatenate the hidden state with the spatio-temporal embedding, and adopt the scaled dot-product approach~\cite{Vaswani-et-al:NIPS2017} to compute the relevance between vertex $ v_i $ and $ v $:
\begin{equation}
s_{v_i,v} = \dfrac{ \langle h_{v_i,t_j}^{(l-1)} \parallel e_{v_i,t_j}, h_{v,t_j}^{(l-1)} \parallel e_{v,t_j} \rangle } {\sqrt{2D}}, 
\end{equation}
where $ \parallel $ represents the concatenation operation, $ \langle \bullet , \bullet \rangle $ denotes the inner product operator, and $ 2D $ is the dimension of $ h_{v_i,t_j}^{(l-1)} \parallel e_{v_i,t_j} $. Then, $ s_{v_i,v} $ is normalized via softmax as:
\begin{equation}
\alpha_{v_i,v} = \dfrac{ \exp ( s_{v_i,v} ) }{ \sum\nolimits_{v_r \in \mathcal{V}} \exp ( s_{v_i,v_r} ) }.
\end{equation}
After the attention score $ \alpha_{v_i,v} $ is obtained, the hidden state can be updated through Equation \ref{spatial attention3}.

\begin{figure}
	\centering
	\includegraphics[width = 0.90 \columnwidth]{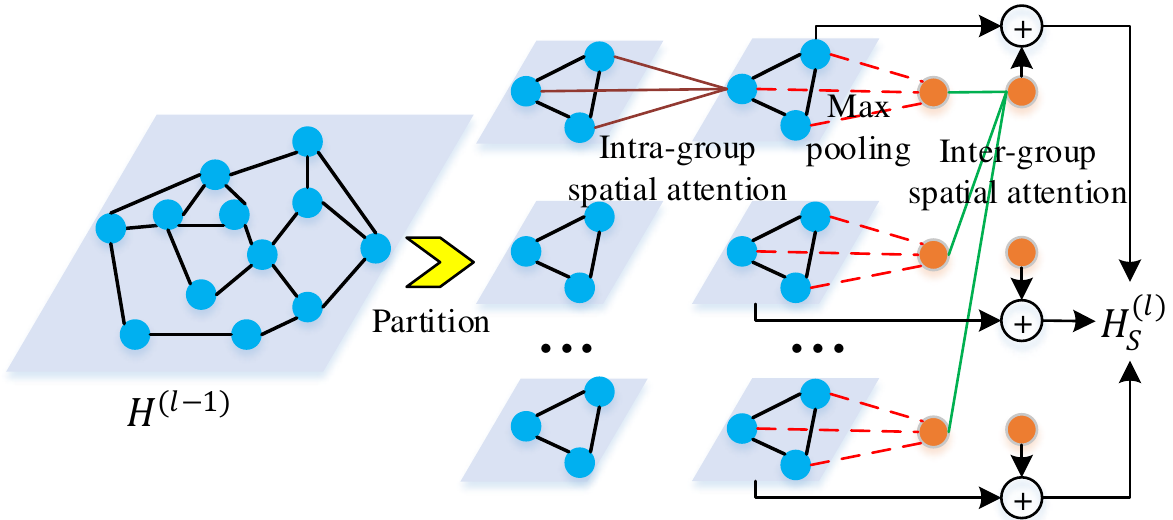} \\
	\caption{Group spatial attention computes both intra-group and inter-group attention to model spatial correlations.}
	\label{Figure4}
\end{figure}

To stabilize the learning process, we extend the spatial attention mechanism to be multi-head ones~\cite{Vaswani-et-al:NIPS2017}. Specifically, we concatenate $ K $ parallel attention mechanisms with different learnable projections:
\begin{equation}
s_{v_i,v}^{(k)} = \dfrac{ \langle f^{(k)}_{s,1} (  h_{v_i,t_j}^{(l-1)} \parallel e_{v_i,t_j} ) , f^{(k)}_{s,2} ( h_{v,t_j}^{(l-1)} \parallel e_{v,t_j} ) \rangle }{\sqrt{d}}
\label{multi-head spatial attention1}, 
\end{equation}
\begin{equation}
\alpha_{v_i,v}^{(k)} = \dfrac{ \exp ( s_{v_i,v}^{(k)} ) }{ \sum_{v_r \in \mathcal{V}} \exp ( s_{v_i,v_r}^{(k)} ) }
\label{multi-head spatial attention2},
\end{equation}
\begin{equation}
hs_{v_i,t_j}^{(l)} = \parallel_{k = 1}^K \left\{ \sum\nolimits_{v \in \mathcal{V}} \alpha_{v_i,v}^{(k)} \cdot f^{(k)}_{s,3} ( h_{v,t_j}^{(l-1)} ) \right\}
\label{multi-head spatial attention3}, 
\end{equation}
where $ f^{(k)}_{s,1} ( \bullet ) $, $ f^{(k)}_{s,2} ( \bullet ) $, and $ f^{(k)}_{s,3} ( \bullet ) $ represent three different nonlinear projections (Equation \ref{non-linear transform}) in the $ k^{th} $ head attention, producing $ d = D / K $ dimensional outputs. 

When the number of vertices $ N $ is large, the time and memory consumption is heavy as we need to compute $ N^2 $ attention scores. To address this limitation, we further propose a group spatial attention, which contains intra-group spatial attention and inter-group spatial attention, as shown in Figure~\ref{Figure4}.

We randomly partition $ N $ vertices into $ G $ groups, where each group contains $ M=N/G $ vertices (padding can be applied if necessary). In each group, we compute the intra-group attention to model the local spatial correlations among vertices through Equations \ref{multi-head spatial attention1}, \ref{multi-head spatial attention2} and \ref{multi-head spatial attention3}, where the learnable parameters are shared across groups. Then, we apply the max-pooling approach in each group to obtain a single representation for each group. Next, we compute the inter-group spatial attention to model the correlations between different groups, producing a global feature for each group. The local feature is added to the corresponding global feature as the final output.

In the group spatial attention, we need to compute $ GM^2+G^2=NM+(N/M)^2 $ attention scores at each time step. By letting the gradient to zero, we know when $ M=\sqrt[3]{2N} $, the number of attention scores reaches its minimum $ 2^{-1/3}N^{4/3} \ll N^2 $.

\begin{figure}
	\centering
	\includegraphics[width = 0.90 \columnwidth]{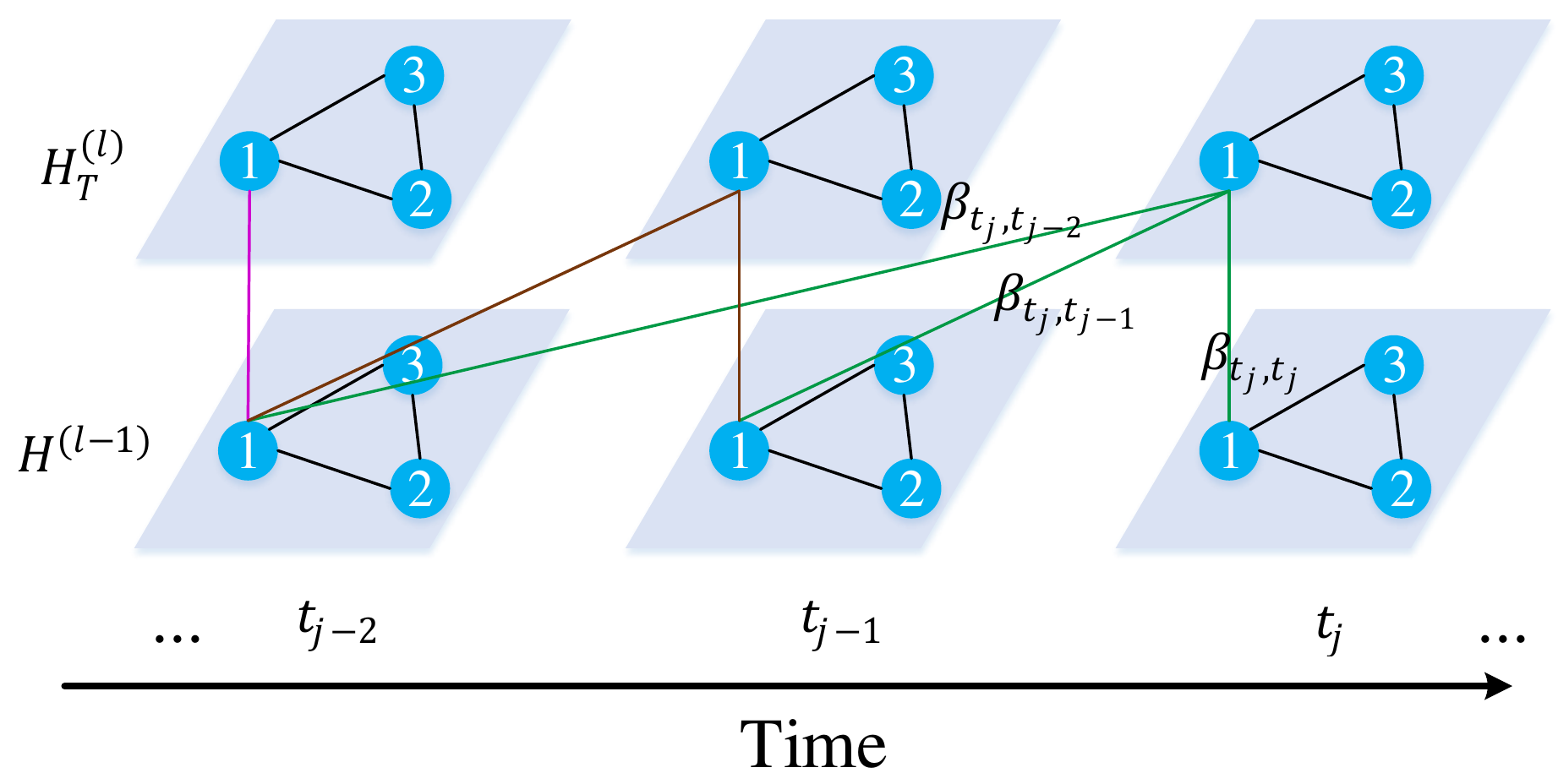} \\
	\caption{The temporal attention mechanism models the non-linear correlations between different time steps.}
	\label{Figure5}
\end{figure}

\subsubsection{Temporal Attention}

The traffic condition at a location is correlated with its previous observations, and the correlations vary over time steps non-linearly. To model these properties, we design a temporal attention mechanism to adaptively model the non-linear correlations between different time steps, as illustrated in Figure \ref{Figure5}. Note that the temporal correlation is influenced by both the traffic conditions and the corresponding time context. For example, a congestion occurring in morning peak hours may affect the traffic for a few hours. Thus, we consider both traffic features and time to measure the relevance between different time steps. Specifically, we concatenate the hidden state with the spatio-temporal embedding, and adopt the multi-head approach to compute the attention score. Formally, considering vertex $ v_i $, the correlation between time step $ t_j $ and $ t $ is defined as:
\begin{equation}
u_{t_j,t}^{(k)} = \dfrac{ \langle f_{t,1}^{(k)} ( h_{v_i,t_j}^{(l-1)} \parallel e_{v_i,t_j} ), f_{t,2}^{(k)} ( h_{v_i,t}^{(l-1)} \parallel e_{v_i,t} ) \rangle }{\sqrt{d}}
\label{multi-head temporal attention1}, 
\end{equation}
\begin{equation}
\beta_{t_j,t}^{(k)} = \dfrac{ \exp ( u_{t_j,t}^{(k)} ) }{ \sum_{t_r \in \mathcal{N}_{t_j}} \exp ( u_{t_j,t_r}^{(k)} ) }
\label{multi-head temporal attention2},
\end{equation}
where $ u_{t_j,t}^{(k)} $ denotes the relevance between time step $ t_j $ and $ t $,  $ \beta_{t_j,t}^{(k)} $ is the attention score in $ k^{th} $ head indicating the importance of time step $ t $ to $ t_j $, $ f_{t,1}^{(k)} ( \bullet ) $ and $ f_{t,2}^{(k)} ( \bullet ) $ represent two different learnable transforms, $ \mathcal{N}_{t_j} $ denotes a set of time steps before $ t_j $, i.e., only considers information from time steps earlier than the target step to enable causality. Once the attention score is obtained, the hidden state of vertex $ v_i $ at time step $ t_j $ is updated as follows:
\begin{equation}
ht_{v_i,t_j}^{(l)} = \parallel_{k = 1}^K \left\{ \sum\nolimits_{t \in \mathcal{N}_{t_j}} \beta_{t_j,t}^{(k)} \cdot f_{t,3}^{(k)} ( h_{v_i,t}^{(l-1)} ) \right\}
\label{multi-head temporal attention3},
\end{equation}
where $ f_{t,3}^{(k)} ( \bullet ) $ represents a non-linear projection. The learnable parameters in Equations \ref{multi-head temporal attention1}, \ref{multi-head temporal attention2} and \ref{multi-head temporal attention3} are shared across all vertices and time steps with paralleled computing.

\begin{figure}
	\centering
	\includegraphics[width = 0.90 \columnwidth]{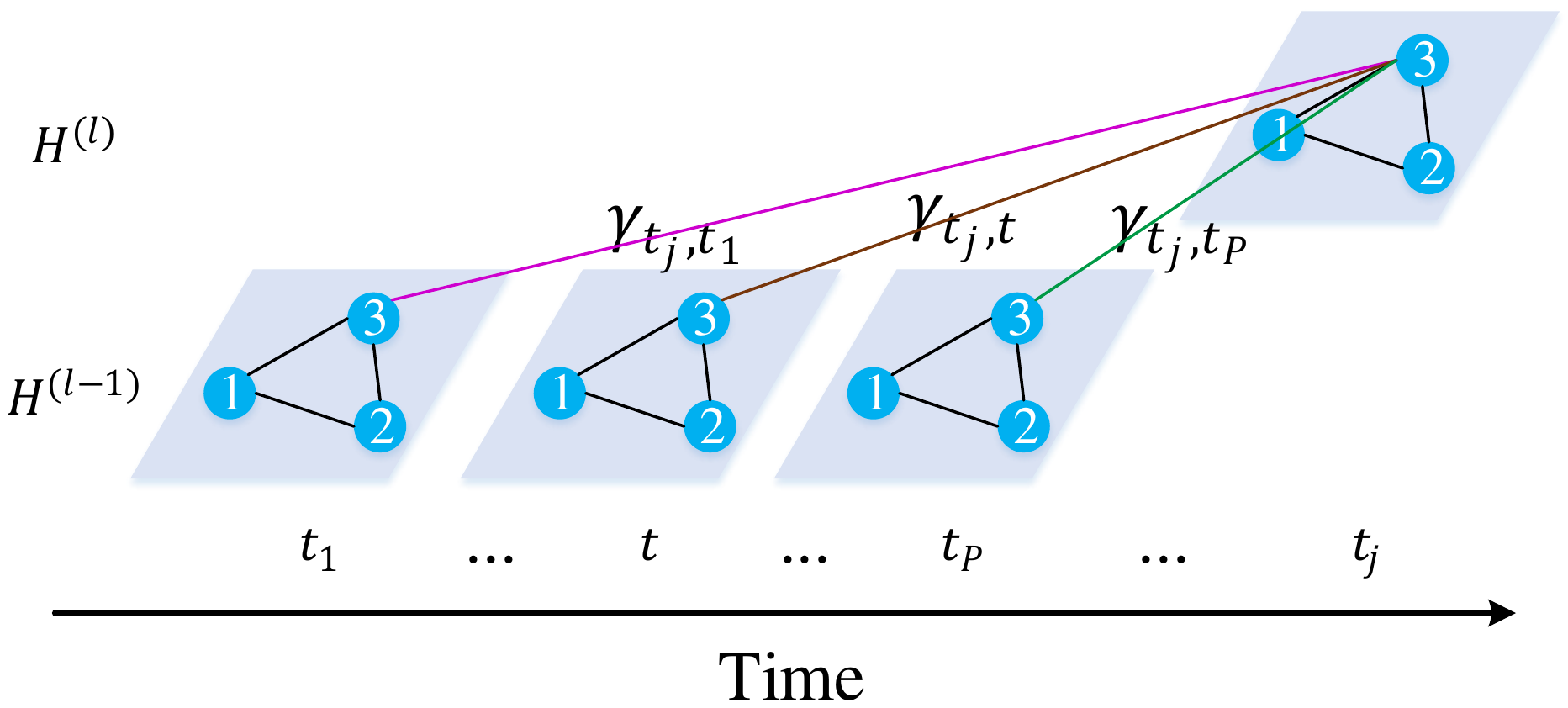} \\
	\caption{The transform attention mechanism models direct relationships between historical  and future time steps.}
	\label{Figure6}
\end{figure}

\subsubsection{Gated Fusion}

The traffic condition of a road at a certain time step is correlated with both its previous values and other roads' traffic conditions. As shown in Figure \ref{Figure2(c)}, we design a gated fusion to adaptively fuse the spatial and temporal representations. In the $ l^{th} $ block, the outputs of the spatial and temporal attention mechanisms are represented as $ H_S^{(l)} $ and $ H_T^{(l)} $, both have the shapes of $ \mathbb{R}^{P \times N \times D} $ in the encoder or $ \mathbb{R}^{Q \times N \times D} $ in the decoder. $ H_S^{(l)} $ and $ H_T^{(l)} $ are fused as:
\begin{equation}
H^{(l)} = z \odot H_S^{(l)} + (1-z) \odot H_T^{(l)},
\end{equation}
with
\begin{equation}
z = \sigma ( H_S^{(l)} \mathbf{W}_{z,1} + H_T^{(l)} \mathbf{W}_{z,2} + \mathbf{b}_z ),
\end{equation}
where $ \mathbf{W}_{z,1} \in \mathbb{R}^{D \times D} $, $ \mathbf{W}_{z,2} \in \mathbb{R}^{D \times D} $ and $ \mathbf{b}_{z} \in \mathbb{R}^{D} $ are learnable parameters, $ \odot $ represents the element-wise product, $ \sigma ( \bullet ) $ denotes the sigmoid activation, $ z $ is the gate. The gated fusion mechanism adaptively controls the flow of spatial and temporal dependencies at each vertex and time step.  

\subsection{Transform Attention}

To ease the error propagation effect between different prediction time steps in the long time horizon, we add a transform attention layer between the encoder and the decoder. It models the direct relationship between each future time step and every historical time step to convert the encoded traffic features to generate future representations as the input of the decoder. As shown in Figure \ref{Figure6}, for vertex $ v_i $, the relevance between the prediction time step $ t_j \ ( t_j = t_{P + 1}, ..., t_{P + Q} ) $ and the historical time step $ t \ ( t = t_1, ..., t_P ) $ is measured via the spatio-temporal embedding:
\begin{equation}
\lambda_{t_j,t}^{(k)} = \dfrac{ \langle f_{tr,1}^{(k)} ( e_{v_i,t_j} ),  f_{tr,2}^{(k)} ( e_{v_i,t} ) \rangle }{\sqrt{d}} 
\label{multi-head transform attention1}, 
\end{equation}
\begin{equation}
\gamma_{t_j,t}^{(k)} = \dfrac{ \exp ( \lambda_{t_j,t}^{(k)} ) }{ \sum_{t_r = t_1}^{t_P} \exp ( \lambda_{t_j,t_r}^{(k)} ) } 
\label{multi-head transform attention2}.
\end{equation}
With the attention score $ \gamma_{t_j,t}^{(k)} $, the encoded traffic feature is transformed to the decoder by adaptively selecting relevant features across all historical $ P $ time steps:
\begin{equation}
h_{v_i,t_j}^{(l)} = \parallel_{k = 1}^K \left\{ \sum\nolimits_{t = t_1}^{t_P} \gamma_{t_j,t}^{(k)} \cdot f_{tr,3}^{(k)} ( h_{v_i,t}^{(l-1)} ) \right\} 
\label{multi-head transform attention3}.
\end{equation}
Equations~\ref{multi-head transform attention1},~\ref{multi-head transform attention2}, and~\ref{multi-head transform attention3} can be computed in parallel across all vertices and time steps, sharing the learnable parameters.

\begin{figure}
	\centering
	\subfigure[Xiamen (95 sensors)]{
		\label{Figure7(a)} 
		\includegraphics[width = 0.4 \columnwidth]{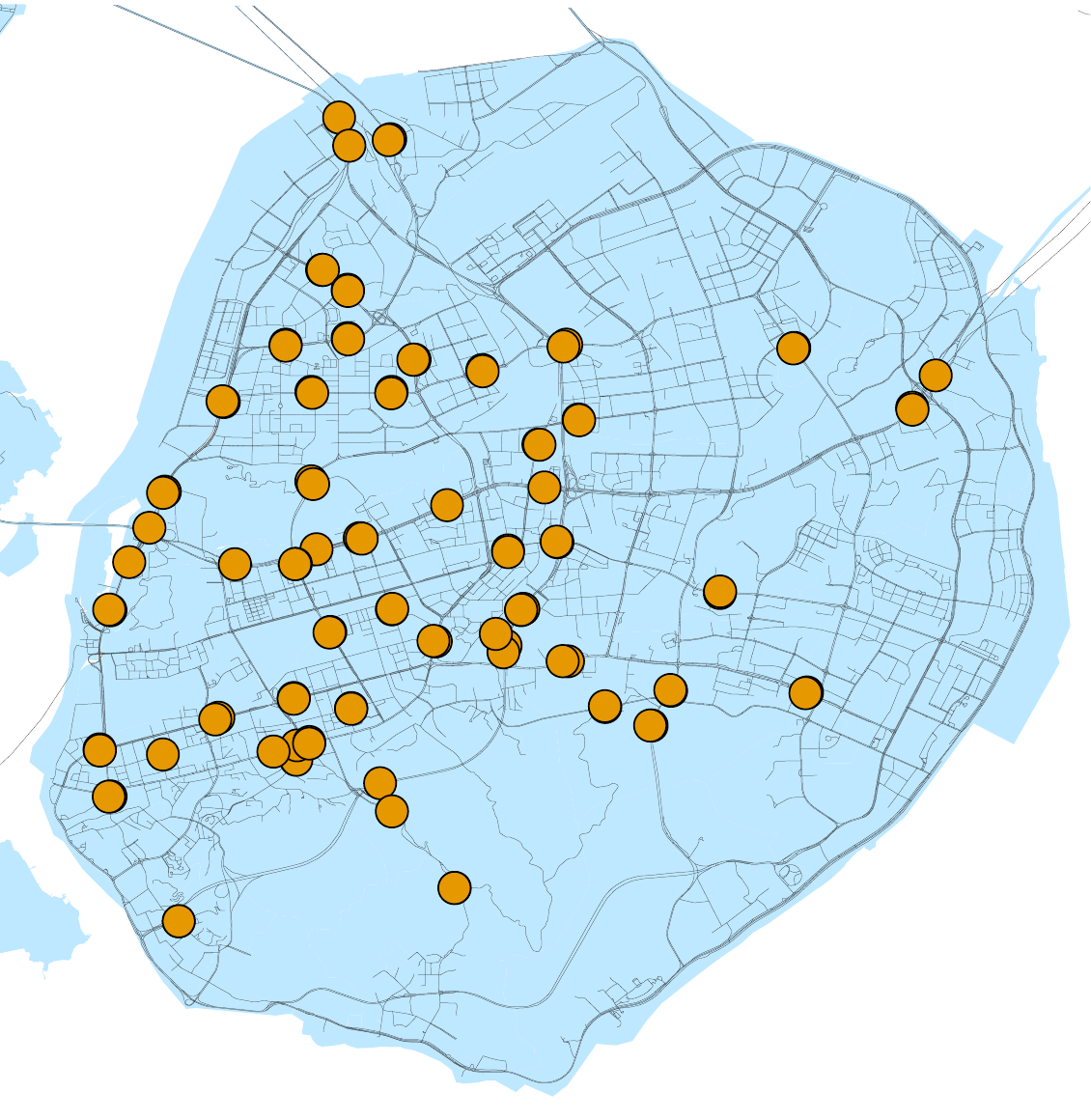}}
	\subfigure[PeMS (325 sensors)]{
		\label{Figure7(b)} 
		\includegraphics[width = 0.5 \columnwidth]{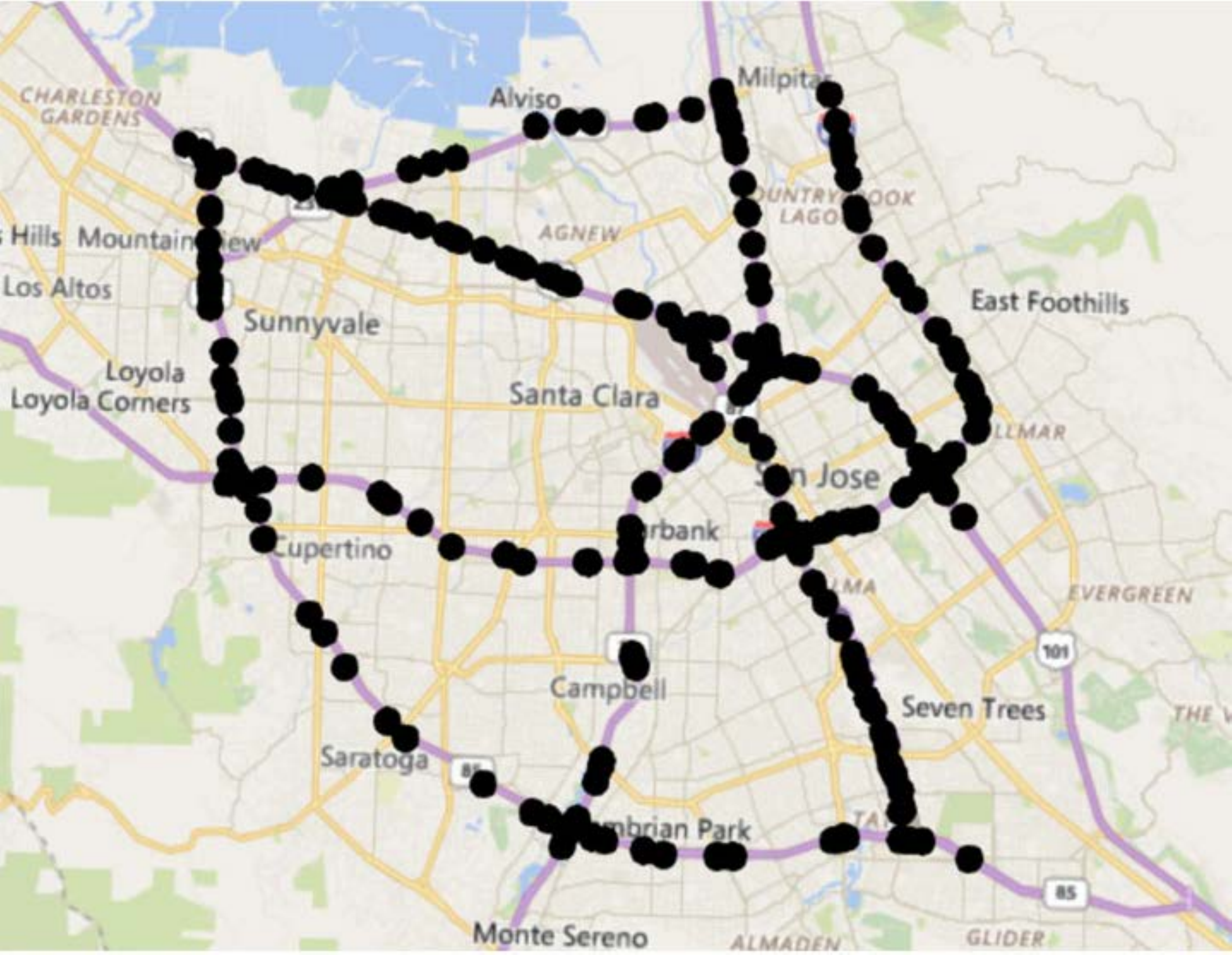}}
	\caption{Sensor distribution of Xiamen and PeMS datasets.}
	\label{Figure7} 
\end{figure}

\begin{table*}
	\centering
	\begin{tabular*}{\hsize}{@{}@{\extracolsep{\fill}}ll|ccc|ccc|ccc@{}}
		\toprule
		\multirow{2}{*}{Data}					& \multirow{2}{*}{Method}	& \multicolumn{3}{c}{15 min}	& \multicolumn{3}{c}{30 min}	& \multicolumn{3}{c}{1 hour}	\\
		&							& MAE	& RMSE	& MAPE			& MAE	& RMSE	& MAPE			& MAE	& RMSE	& MAPE 			\\
		\midrule
		\multirow{8}{*}{\rotatebox{90}{Xiamen}} & ARIMA          			& 14.81	& 25.03	& 18.05\%		& 18.83	& 33.09	& 22.19\%		& 26.58	& 46.32	& 30.76\%		\\
		& SVR                       & 13.05 & 21.47 & 16.46\%       & 15.66 & 26.34 & 19.68\%       & 20.69 & 35.86 & 26.24\%       \\
		& FNN                       & 13.55 & 22.47 & 16.72\%       & 16.80 & 28.71 & 19.97\%       & 22.90 & 39.51 & 26.19\%       \\
		& FC-LSTM       			& 12.51 & 20.79 & 16.08\%       & 13.74 & 23.93 & 17.23\%       & 16.02 & 29.57 & 19.33\%       \\
		& STGCN                     & 11.76 & 19.94 & 14.93\%       & 13.19 & 23.29 & 16.36\%       & 15.83 & 29.40 & 18.66\%       \\
		& DCRNN                     & 11.67 & \textbf{19.40} & 14.85\%       & 12.76 & 22.20 & 15.99\%       & 14.30 & 25.86 & 17.17\%       \\
		& Graph WaveNet             & \textbf{11.26} & 19.57 & \textbf{14.39\%}       & 12.06 & 21.61 & 15.39\%       & 13.33 & 24.77 & 16.50\%		\\
		& GMAN                      & 11.50 & 19.52 & 14.59\%       & \textbf{12.02} & \textbf{21.42} & \textbf{15.14\%}       & \textbf{12.79} & \textbf{24.15} & \textbf{15.84\%}		\\
		\midrule	
		\multirow{8}{*}{\rotatebox{90}{PeMS}} 	& ARIMA          			& 1.62	& 3.30	& 3.50\%		& 2.33	& 4.76	& 5.40\%		& 3.38	& 6.50	& 8.30\%		\\
		& SVR                       & 1.85  & 3.59  & 3.80\%        & 2.48  & 5.18  & 5.50\%        & 3.28  & 7.08  & 8.00\%		\\
		& FNN                       & 2.20  & 4.42  & 5.19\%        & 2.30  & 4.63  & 5.43\%        & 2.46  & 4.98  & 5.89\%        \\
		& FC-LSTM       			& 2.05  & 4.19  & 4.80\%        & 2.20  & 4.55  & 5.20\%        & 2.37  & 4.96  & 5.70\%        \\
		& STGCN                     & 1.36  & 2.96  & 2.90\%        & 1.81  & 4.27  & 4.17\%        & 2.49  & 5.69  & 5.79\%        \\
		& DCRNN                     & 1.38  & 2.95  & 2.90\%        & 1.74  & 3.97  & 3.90\%        & 2.07  & 4.74  & 4.90\%        \\
		& Graph WaveNet             & \textbf{1.30}  & \textbf{2.74}	& \textbf{2.73\%}		& 1.63	& \textbf{3.70}	& 3.67\%		& 1.95	& 4.52	& 4.63\%		\\
		& GMAN                      & 1.34	& 2.82	& 2.81\%		& \textbf{1.62}	& 3.72	& \textbf{3.63\%}		& \textbf{1.86}	& \textbf{4.32}	& \textbf{4.31\%}		\\		
		\bottomrule
	\end{tabular*}
	\caption{Performance comparison of different approaches for traffic prediction on Xiamen and PeMS datasets.}
	\label{Table1}
\end{table*}

\subsection{Encoder-Decoder}

As shown in Figure \ref{Figure2(a)}, GMAN is an encoder-decoder architecture. Before entering into the encoder, the historical observation $ \mathcal{X} \in \mathbb{R}^{P \times N \times C} $ is transformed to $ H^{(0)} \in \mathbb{R}^{P \times N \times D} $ using fully-connected layers. Then, $ H^{(0)} $ is fed into the encoder with $ L $ ST-Attention blocks, and produces an output $ H^{(L)} \in \mathbb{R}^{P \times N \times D} $. Following the encoder, a transform attention layer is added to convert the encoded feature $ H^{(L)} $ to generate the future sequence representation $ H^{(L+1)} \in \mathbb{R}^{Q \times N \times D} $. Next, the decoder stacks $ L $ ST-Attention blocks upon $ H^{(L+1)} $, and produces the output as $ H^{(2L+1)} \in \mathbb{R}^{Q \times N \times D} $. Finally, the fully-connected layers produce the $ Q $ time steps ahead prediction $ \hat{Y} \in \mathbb{R}^{Q \times N \times C} $.

GMAN can be trained end-to-end via back-propagation by minimizing the \textit{mean absolute error} (MAE) between predicted values and ground truths:
\begin{equation}
\mathcal{L} (\Theta) = \frac{1}{Q} \sum\nolimits_{t = t_{P + 1}}^{t_{P + Q}} \left| Y_{t} - \hat{Y}_{t} \right|, 
\end{equation}
where $ \Theta $ denotes all learnable parameters in GMAN.

\section{Experiments}

\subsection{Datasets} \label{datasets}

We evaluate the performance of GMAN on two traffic prediction tasks with different road network scales: (1) traffic volume prediction on the \textbf{Xiamen} dataset~\cite{Wang-et-al:ICWS2017}, which contains 5 months of data recorded by 95 traffic sensors ranging from August 1st, 2015 to December 31st, 2015 in Xiamen, China; (2) traffic speed prediction on the \textbf{PeMS} dataset~\cite{Li-et-al:ICLR2018}), which contains 6 months of data recorded by 325 traffic sensors ranging from January 1st, 2017 to June 30th, 2017 in the Bay Area. The distributions of sensors in two datasets are visualized in Figure \ref{Figure7}.

\paragraph{Data Preprocessing} 
We adopt the same data preprocessing procedures as in~\cite{Li-et-al:ICLR2018}. In both datasets, a time step denotes 5 minutes and the data is normalized via the Z-Score method. We use 70\% of the data for training, 10\% for validation, and 20\% for testing. To construct the road network graph, each traffic sensor is considered as a vertex and we compute the pairwise road network distances between sensors. Then, the adjacency matrix is defined as:
\begin{equation}
\mathcal{A}_{v_i, v_j} = \left\{ 
\begin{array}{lr}
\exp ( -\dfrac{d_{v_i,v_j}^2}{\sigma^2} ), if \exp ( -\dfrac{d_{v_i,v_j}^2}{\sigma^2} ) \geq \epsilon \\
0, otherwise 
\end{array}
\label{adjacency_matrix},
\right.
\end{equation}
where $ {d_{v_i,v_j}} $ is the road network distance from sensor $ v_i $ to $ v_j $, $ \sigma $ is the standard deviation, and $ \epsilon $ (assigned to 0.1) is the threshold to control the sparsity of the adjacency matrix $ \mathcal{A} $.

\subsection{Experimental Settings}

\paragraph{Metrics} 

We apply three widely used metrics to evaluate the performance of our model, i.e., \textit{Mean Absolute Error} (MAE), \textit{Root Mean Squared Error} (RMSE), and \textit{Mean Absolute Percentage Error} (MAPE).

\paragraph{Hyperparameters}

Following the previous works~\cite{Li-et-al:ICLR2018,Wu-et-al:IJCAI2019}, we use $ P=12 $ historical time steps (1 hour) to predict the traffic conditions of the next $ Q=12 $ steps (1 hour). We train our model using Adam optimizer~\cite{Kingma-and-Ba:ICLR2015} with an initial learning rate of 0.001. In the group spatial attention, we partition the vertices into $ G=19 $ groups in the Xiamen dataset and $ G=37 $ groups in the PeMS dataset, respectively. The number of traffic conditions on both datasets is $ C=1 $. Totally, there are 3 hyperparameters in our model, i.e., the number of ST-Attention blocks $ L $, the number of attention heads $ K $, and the dimensionality $ d $ of each attention head (the channel of each layer $ D = K \times d $). We tune these parameters on the validation set, and observe the best performance on the setting $ L=3 $, $ K=8 $, and $ d=8 $ ($ D=64 $).

\paragraph{Baselines}

We compare GMAN with the following baseline methods: (1) Auto-regressive integrated moving average (\textbf{ARIMA})~\cite{ARIMA:1997}; (2) Support vector regression (\textbf{SVR})~\cite{Wu-et-al:TITS2004}; (3) Feedforward neural network (\textbf{FNN}); (4) \textbf{FC-LSTM}~\cite{Sutskever-et-al:NIPS2014}, which is a sequence-to-sequence model with fully-connected LSTM layers in both encoder and decoder; (5) Spatio-temporal graph convolutional network (\textbf{STGCN})~\cite{Yu-et-al:IJCAI2018} that combines graph convolutional layers and convolutional sequence learning layers; (6) Diffusion convolutional recurrent neural network (\textbf{DCRNN})~\cite{Li-et-al:ICLR2018} that integrates diffusion convolution with sequence-to-sequence architecture; (7)~\textbf{Graph WaveNet}~\cite{Wu-et-al:IJCAI2019} that combines graph convolution with dilated casual convolution.

For models ARIMA, SVR, FNN, and FC-LSTM, we use the settings suggested by~\cite{Li-et-al:ICLR2018}. For models STGCN, DCRNN, and Graph WaveNet, we use the default settings from their original proposals.

\subsection{Experimental Results}

\subsubsection{Forecasting Performance Comparison}

Table~\ref{Table1} shows the comparison of different methods for 15 minutes (3 steps), 30 minutes (6 steps), and 1 hour (12 steps) ahead predictions on two datasets. We observe that: (1) deep learning approaches outperform traditional time series methods and machine learning models, demonstrating the ability of deep neural networks in modeling non-linear traffic data; (2) among deep learning methods, graph-based models including STGCN, DCRNN, Graph WaveNet, and GMAN generally perform better than FC-LSTM, indicating the road network information is essential for traffic prediction; and (3) GMAN achieves state-of-the-art prediction performances and the advantages are more evident in the long-term horizon (e.g., 1 hour ahead). We argue that the long-term traffic prediction is more beneficial to practical applications, e.g., it allows transportation agencies to have more time to take actions to optimize the traffic according to the prediction.

We also use the \textit{T-Test} to test the significance of GMAN in 1 hour ahead prediction compared to Graph WaveNet. The p-value is less than 0.01, which demonstrates that GMAN statistically outperforms Graph WaveNet.

\subsubsection{Fault Tolerance Comparison}

\begin{figure}
	\centering
	\subfigure[Xiamen]{
		\label{Figure8(a)} 
		\includegraphics[width = 0.45 \columnwidth]{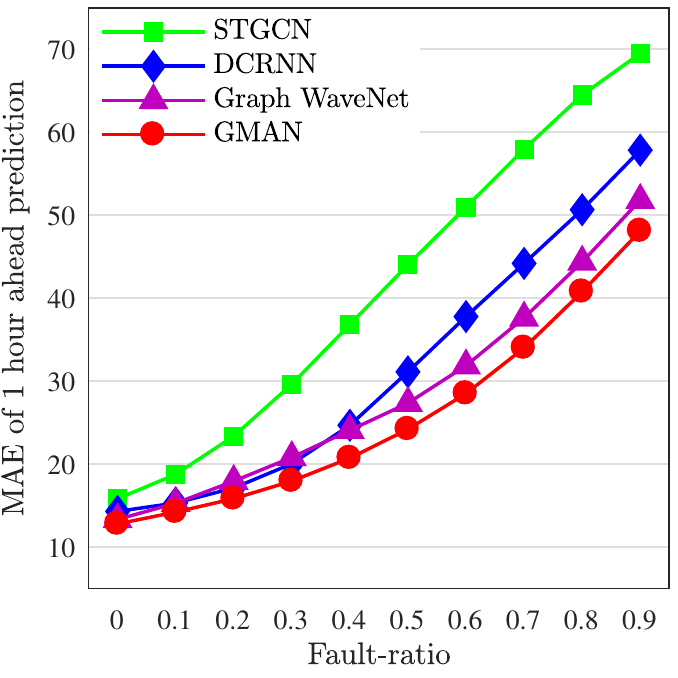}}
	\subfigure[PeMS]{
		\label{Figure8(b)} 
		\includegraphics[width = 0.45 \columnwidth]{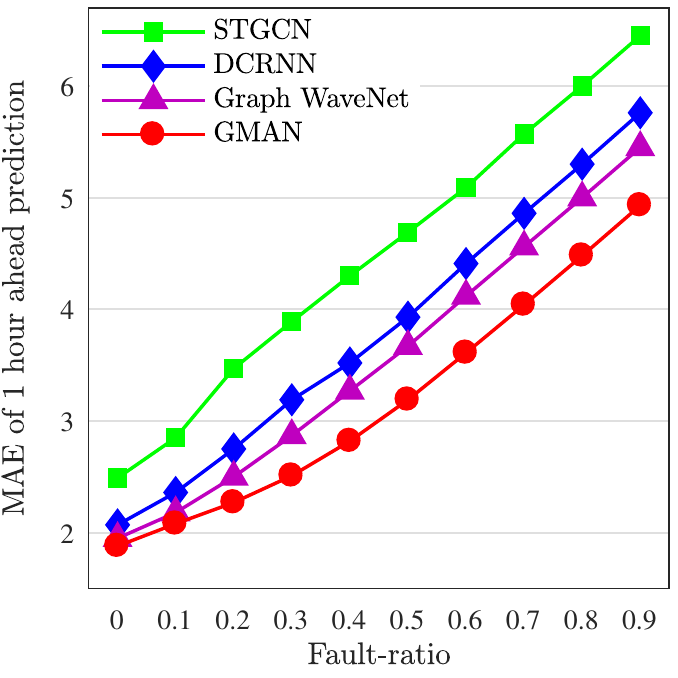}}
	\caption{Fault-tolerance comparison.}
	\label{Figure8} 
\end{figure}

The real-time values of traffic conditions may be missing partially, due to sensor malfunction, packet losses during data transmission, etc. To evaluate the fault-tolerance ability, we randomly drop a fraction $ \eta $ (fault-ratio, ranging from 10\% to 90\%) of historical observations (i.e., randomly replace $ \eta \times N \times P \times C $ input values with zeros) to make 1 hour ahead predictions. As shown in Figure~\ref{Figure8}, GMAN is more fault tolerant than state-of-the-art methods. This shows that GMAN can capture the complex spatio-temporal correlations from the ``contaminated" traffic data and adjust the dependencies from observations to future time steps.

\subsubsection{Effect of Each Component}

\begin{figure}
	\centering
	\subfigure[Xiamen]{
		\label{Figure9(a)} 
		\includegraphics[width = 0.45 \columnwidth]{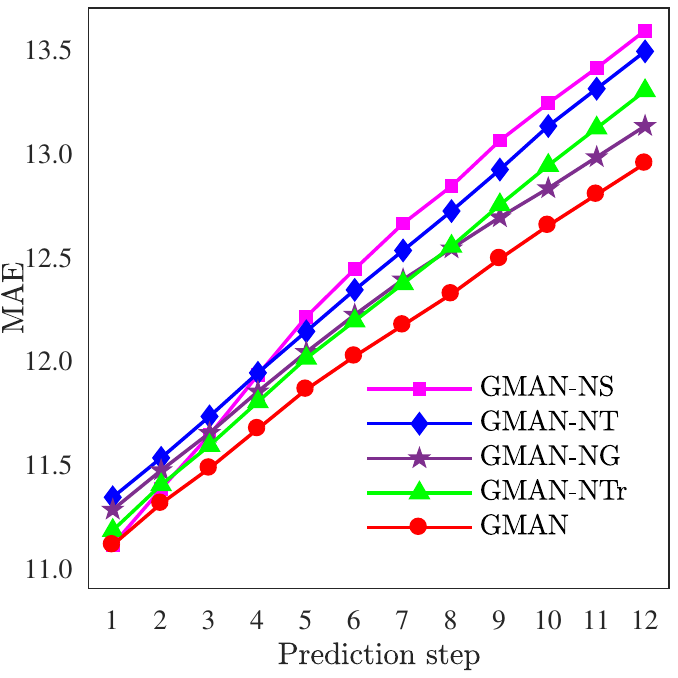}}
	\subfigure[PeMS]{
		\label{Figure9(b)} 
		\includegraphics[width = 0.45 \columnwidth]{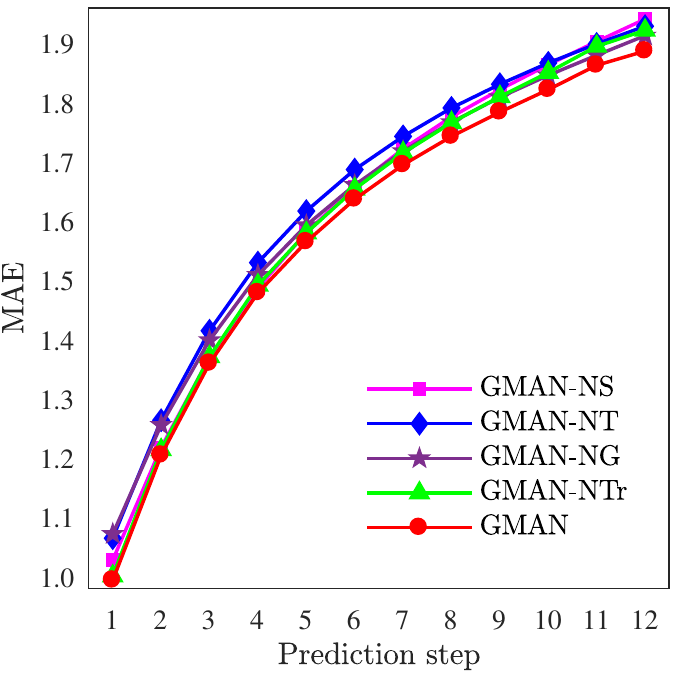}}
	\caption{MAE of each prediction step.}
	\label{Figure9} 
\end{figure}

To investigate the effect of each component in our model, we evaluate four variants by removing spatial attention, temporal attention, gated fusion, and transform attention from GMAN separately, which are named as GMAN-NS, GMAN-NT, GMAN-NG, and GMAN-NTr respectively. Figure~\ref{Figure9} presents the MAE in each prediction step of GMAN and the four variants. We observe that GMAN consistently outperforms GMAN-NS, GMAN-NT, and GMAN-NG, indicating the effectiveness of spatial attention, temporal attention, and gated fusion in modeling the complex spatio-temporal correlations. Moreover, GMAN performs better than GMAN-NTr, especially in the long-term horizon, demonstrating that the transform attention mechanism effectively eases the effect of error propagation.

\subsubsection{Computation Time}

We present the training time and inference time of STGCN, DCRNN, Graph WaveNet, and GMAN on the PeMS dataset in Table~\ref{Table2}. During the training phase, GMAN has a similar speed with Graph WaveNet. DCRNN runs much slower than other methods due to the time-consuming sequence learning in recurrent networks. STGCN is the most efficient but shows poor prediction performance (Table~\ref{Table1}). In the inference phase, we report the total time cost on the validation data. STGCN and DCRNN is less efficient as they need iterative computation to generate the 12 prediction results. GMAN and Graph WaveNet could produce 12 steps ahead predictions in one run and thus take less time for inference. 

In respect of the second best model Graph WaveNet as suggested in Table~\ref{Table1}, GMAN compares favorably to Graph WaveNet in the long-term (e.g., 1 hour ahead) traffic predictions (Table~\ref{Table1}) with similar computation costs for both training and inference (Table~\ref{Table2}). 

\begin{table}
	\centering
	\begin{tabular*}{\hsize}{@{}@{\extracolsep{\fill}}lrr@{}}
		\toprule
		\multirow{2}{*}{Method}		& \multicolumn{2}{c}{Computation Time}		\\
		& Training (s/epoch)	& Inference (s)		\\
		\midrule
		STGCN						& 51.35					& 94.56 			\\
		DCRNN						& 650.64				& 110.52			\\
		Graph WaveNet				& 182.21				& 6.55				\\
		GMAN                        & 217.62                & 9.34              \\
		\bottomrule
	\end{tabular*}
	\caption{The computation time on the PeMS dataset.}
	\label{Table2}
\end{table}

\section{Conclusion}

We proposed a graph multi-attention network (GMAN) to predict traffic conditions for time steps ahead on a road network graph. Specifically, we proposed spatial and temporal attention mechanisms with gated fusion to model the complex spatio-temporal correlations. We further designed a transform attention mechanism to ease the effect of error propagation to improve the long-term prediction performance. Experiments on two real-world datasets show that GMAN achieves state-of-the-art results, and the advantages are more evident as the predictions are made into far future. In the future, we will apply GMAN to other spatio-temporal prediction tasks, such as water consumption prediction. 

\section{Acknowledgments}

This work was supported by grants from Natural Science Foundation of China (61872306 and U1605254), and Xiamen Science and Technology Bureau (3502Z20193017).

\bibliographystyle{aaai}
\bibliography{AAAI-ZhengZ.2220}

\begin{thebibliography}{}

\bibitem[\protect\citeauthoryear{Atwood and
  Towsley}{2016}]{Atwood-and-Towsley:NIPS2016}
Atwood, J., and Towsley, D.
\newblock 2016.
\newblock Diffusion-convolutional neural networks.
\newblock In {\em NeurIPS},  1993--2001.

\bibitem[\protect\citeauthoryear{Bronstein \bgroup et al\mbox.\egroup
  }{2017}]{Bronstein-et-al:SPM2017}
Bronstein, M.~M.; Bruna, J.; LeCun, Y.; Szlam, A.; and Vandergheynst, P.
\newblock 2017.
\newblock Geometric deep learning: going beyond euclidean data.
\newblock {\em IEEE Signal Processing Magazine} 34(4):18--42.

\bibitem[\protect\citeauthoryear{Chen, Ma, and
  Xiao}{2018}]{Chen-et-al:ICLR2018}
Chen, J.; Ma, T.; and Xiao, C.
\newblock 2018.
\newblock Fastgcn: Fast learning with graph convolutional networks via
  importance sampling.
\newblock In {\em ICLR}.

\bibitem[\protect\citeauthoryear{Cheng \bgroup et al\mbox.\egroup
  }{2018}]{Cheng-et-al:AAAI2018}
Cheng, W.; Shen, Y.; Zhu, Y.; and Huang, L.
\newblock 2018.
\newblock A neural attention model for urban air quality inference: learning
  the weights of monitoring stations.
\newblock In {\em AAAI},  2151--2158.

\bibitem[\protect\citeauthoryear{Cui \bgroup et al\mbox.\egroup
  }{2019}]{Cui-et-al:TKDE2019}
Cui, P.; Wang, X.; Pei, J.; and Zhu, W.
\newblock 2019.
\newblock A survey on network embedding.
\newblock {\em IEEE Transactions on Knowledge and Data Engineering}
  31(5):833--852.

\bibitem[\protect\citeauthoryear{Defferrard, Bresson, and
  Vandergheynst}{2016}]{Defferrard-et-al:NIPS2016}
Defferrard, M.; Bresson, X.; and Vandergheynst, P.
\newblock 2016.
\newblock Convolutional neural networks on graphs with fast localized spectral
  filtering.
\newblock In {\em NeurIPS},  3844--3852.

\bibitem[\protect\citeauthoryear{Du \bgroup et al\mbox.\egroup
  }{2018}]{Du-et-al:arXiv2018}
Du, S.; Li, T.; Gong, X.; and Horng, S.-J.
\newblock 2018.
\newblock A hybrid method for traffic flow forecasting using multimodal deep
  learning.
\newblock {\em arXiv preprint arXiv:1803.02099}.

\bibitem[\protect\citeauthoryear{Grover and
  Leskovec}{2016}]{Grover-and-Leskovec:KDD2016}
Grover, A., and Leskovec, J.
\newblock 2016.
\newblock Node2vec: scalable feature learning for networks.
\newblock In {\em KDD},  855--864.

\bibitem[\protect\citeauthoryear{Hamilton, Ying, and
  Leskovec}{2017}]{Hamilton-et-al:NIPS2017}
Hamilton, W.~L.; Ying, R.; and Leskovec, J.
\newblock 2017.
\newblock Inductive representation learning on large graphs.
\newblock In {\em NeurIPS},  1024--1034.

\bibitem[\protect\citeauthoryear{He \bgroup et al\mbox.\egroup
  }{2016}]{He-et-al:CVPR2016}
He, K.; Zhang, X.; Ren, S.; and Sun, J.
\newblock 2016.
\newblock Deep residual learning for image recognition.
\newblock In {\em CVPR},  770--778.

\bibitem[\protect\citeauthoryear{Hou and Li}{2016}]{Hou-and-Li:TITS2016}
Hou, Z., and Li, X.
\newblock 2016.
\newblock Repeatability and similarity of freeway traffic flow and long-term
  prediction under big data.
\newblock {\em IEEE Transactions on Intelligent Transportation Systems}
  17(6):1786--1796.

\bibitem[\protect\citeauthoryear{Kingma and Ba}{2015}]{Kingma-and-Ba:ICLR2015}
Kingma, D.~P., and Ba, J.~L.
\newblock 2015.
\newblock Adam: a method for stochastic optimization.
\newblock In {\em ICLR}.

\bibitem[\protect\citeauthoryear{Kipf and
  Welling}{2017}]{Kipf-and-Welling:ICLR2017}
Kipf, T.~N., and Welling, M.
\newblock 2017.
\newblock Semi-supervised classification with graph convolutional networks.
\newblock In {\em ICLR}.

\bibitem[\protect\citeauthoryear{Li \bgroup et al\mbox.\egroup
  }{2018a}]{Li-et-al:AAAI2018}
Li, R.; Wang, S.; Zhu, F.; and Huang, J.
\newblock 2018a.
\newblock Adaptive graph convolutional neural networks.
\newblock In {\em AAAI}.

\bibitem[\protect\citeauthoryear{Li \bgroup et al\mbox.\egroup
  }{2018b}]{Li-et-al:ICLR2018}
Li, Y.; Yu, R.; Shahabi, C.; and Liu, Y.
\newblock 2018b.
\newblock Diffusion convolutional recurrent neural network: Data-driven traffic
  forecasting.
\newblock In {\em ICLR}.

\bibitem[\protect\citeauthoryear{Lv \bgroup et al\mbox.\egroup
  }{2018}]{Lv-et-al:IJCAI2018}
Lv, Z.; Xu, J.; Zheng, K.; Yin, H.; Zhao, P.; and Zhou, X.
\newblock 2018.
\newblock Lc-rnn: a deep learning model for traffic speed prediction.
\newblock In {\em IJCAI},  3470--3476.

\bibitem[\protect\citeauthoryear{Ma \bgroup et al\mbox.\egroup
  }{2015}]{Ma-et-al:TRC2015}
Ma, X.; Tao, Z.; Wang, Y.; Yu, H.; and Wang, Y.
\newblock 2015.
\newblock Long short-term memory neural network for traffic speed prediction
  using remote microwave sensor data.
\newblock {\em Transportation Research Part C: Emerging Technologies}
  54:187--197.

\bibitem[\protect\citeauthoryear{Makridakis and Hibon}{1997}]{ARIMA:1997}
Makridakis, S., and Hibon, M.
\newblock 1997.
\newblock Arma models and the box–jenkins methodology.
\newblock {\em Journal of Forecasting} 16(3):147--163.

\bibitem[\protect\citeauthoryear{Nair and
  Hinton}{2010}]{Nair-and-Hinton:ICML2010}
Nair, V., and Hinton, G.~E.
\newblock 2010.
\newblock Rectified linear units improve restricted boltzmann machines.
\newblock In {\em ICML},  807--814.

\bibitem[\protect\citeauthoryear{Shen \bgroup et al\mbox.\egroup
  }{2018}]{Shen-et-al:AAAI2018}
Shen, T.; Jiang, J.; Zhou, T.; Pan, S.; Long, G.; and Zhang, C.
\newblock 2018.
\newblock Disan: Directional self-attention network for rnn/cnn-free language
  understanding.
\newblock In {\em AAAI},  5446--5455.

\bibitem[\protect\citeauthoryear{Song, Kanasugi, and
  Shibasaki}{2016}]{Song-et-al:IJCAI2016}
Song, X.; Kanasugi, H.; and Shibasaki, R.
\newblock 2016.
\newblock Deeptransport: Prediction and simulation of human mobility and
  transportation mode at a citywide level.
\newblock In {\em IJCAI},  2618--2624.

\bibitem[\protect\citeauthoryear{Sutskever, Vinyals, and
  Le}{2014}]{Sutskever-et-al:NIPS2014}
Sutskever, I.; Vinyals, O.; and Le, Q.~V.
\newblock 2014.
\newblock Sequence to sequence learning with neural networks.
\newblock In {\em NeurIPS},  3104--3112.

\bibitem[\protect\citeauthoryear{van~den Oord \bgroup et al\mbox.\egroup
  }{2016}]{Oord-et-al:arXiv2016}
van~den Oord, A.; Dieleman, S.; Zen, H.; Simonyan, K.; Vinyals, O.; Graves, A.;
  Kalchbrenner, N.; Senior, A.; and Kavukcuoglu, K.
\newblock 2016.
\newblock Wavenet: A generative model for raw audio.
\newblock {\em arXiv preprint arXiv:1609.03499}.

\bibitem[\protect\citeauthoryear{Vaswani \bgroup et al\mbox.\egroup
  }{2017}]{Vaswani-et-al:NIPS2017}
Vaswani, A.; Shazeer, N.; Parmar, N.; Uszkoreit, J.; Jones, L.; Gomez, A.~N.;
  Łukasz Kaiser; and Polosukhin, I.
\newblock 2017.
\newblock Attention is all you need.
\newblock In {\em NeurIPS},  5998--6008.

\bibitem[\protect\citeauthoryear{Veličković \bgroup et al\mbox.\egroup
  }{2018}]{Velickovic-et-al:ICLR2018}
Veličković, P.; Cucurull, G.; Casanova, A.; Romero, A.; Liò, P.; and Bengio,
  Y.
\newblock 2018.
\newblock Graph attention networks.
\newblock In {\em ICLR}.

\bibitem[\protect\citeauthoryear{Wang \bgroup et al\mbox.\egroup
  }{2017}]{Wang-et-al:ICWS2017}
Wang, Y.; Fan, X.; Liu, X.; Zheng, C.; Chen, L.; Wang, C.; and Li, J.
\newblock 2017.
\newblock Unlicensed taxis detection service based on large-scale vehicles
  mobility data.
\newblock In {\em ICWS},  857--861.

\bibitem[\protect\citeauthoryear{Wu \bgroup et al\mbox.\egroup
  }{2019a}]{Wu-et-al:arXiv2019}
Wu, Z.; Pan, S.; Chen, F.; Long, G.; Zhang, C.; and Yu, P.~S.
\newblock 2019a.
\newblock A comprehensive survey on graph neural networks.
\newblock {\em arXiv preprint arXiv:1901.00596}.

\bibitem[\protect\citeauthoryear{Wu \bgroup et al\mbox.\egroup
  }{2019b}]{Wu-et-al:IJCAI2019}
Wu, Z.; Pan, S.; Long, G.; Jiang, J.; and Zhang, C.
\newblock 2019b.
\newblock Graph wavenet for deep spatial-temporal graph modeling.
\newblock In {\em IJCAI}.

\bibitem[\protect\citeauthoryear{Wu, Ho, and Lee}{2004}]{Wu-et-al:TITS2004}
Wu, C.-H.; Ho, J.-M.; and Lee, D.~T.
\newblock 2004.
\newblock Travel-time prediction with support vector regression.
\newblock {\em IEEE Transactions on Intelligent Transportation Systems}
  5(4):276--281.

\bibitem[\protect\citeauthoryear{Yao \bgroup et al\mbox.\egroup
  }{2018}]{Yao-et-al:AAAI2018}
Yao, H.; Wu, F.; Ke, J.; Tang, X.; Jia, Y.; Lu, S.; Gong, P.; Ye, J.; and Li,
  Z.
\newblock 2018.
\newblock Deep multi-view spatial-temporal network for taxi demand prediction.
\newblock In {\em AAAI},  2588--2595.

\bibitem[\protect\citeauthoryear{Yao \bgroup et al\mbox.\egroup
  }{2019}]{Yao-et-al:AAAI2019}
Yao, H.; Tang, X.; Wei, H.; Zheng, G.; and Li, Z.
\newblock 2019.
\newblock Revisiting spatial-temporal similarity: A deep learning framework for
  traffic prediction.
\newblock In {\em AAAI}.

\bibitem[\protect\citeauthoryear{Yu, Yin, and Zhu}{2018}]{Yu-et-al:IJCAI2018}
Yu, B.; Yin, H.; and Zhu, Z.
\newblock 2018.
\newblock Spatio-temporal graph convolutional networks: A deep learning
  framework for traffic forecasting.
\newblock In {\em IJCAI},  3634--3640.

\bibitem[\protect\citeauthoryear{Zhang, Zheng, and
  Qi}{2017}]{Zhang-et-al:AAAI2017}
Zhang, J.; Zheng, Y.; and Qi, D.
\newblock 2017.
\newblock Deep spatio-temporal residual networks for citywide crowd flows
  prediction.
\newblock In {\em AAAI},  1655--1661.

\bibitem[\protect\citeauthoryear{Zheng and Su}{2014}]{Zheng-et-al:TRC2014}
Zheng, Z., and Su, D.
\newblock 2014.
\newblock Short-term traffic volume forecasting: A k-nearest neighbor approach
  enhanced by constrained linearly sewing principle component algorithm.
\newblock {\em Transportation Research Part C: Emerging Technologies}
  43:143--157.

\bibitem[\protect\citeauthoryear{Zheng \bgroup et al\mbox.\egroup
  }{2019}]{Zheng-et-al:TITS2019}
Zheng, C.; Fan, X.; Wen, C.; Chen, L.; Wang, C.; and Li, J.
\newblock 2019.
\newblock Deepstd: Mining spatio-temporal disturbances of multiple context
  factors for citywide traffic flow prediction.
\newblock {\em IEEE Transactions on Intelligent Transportation Systems}.
\newblock to be published.

\end{thebibliography}

\end{document}